\documentclass[fleqn,usenatbib]{mnras}
\usepackage{newtxtext,newtxmath,amsmath,bm,siunitx}
\usepackage{graphicx}
\usepackage[T1]{fontenc}
\DeclareRobustCommand{\VAN}[3]{#2}
\let\VANthebibliography\thebibliography
\def\thebibliography{\DeclareRobustCommand{\VAN}[3]{##3}\VANthebibliography}

\newcommand{\Alfven}{Alfv$\acute{\text{e}}$n\,} 

\usepackage{lineno}
\newcommand*\patchAmsMathEnvironmentForLineno[1]{
  \expandafter\let\csname old#1\expandafter\endcsname\csname #1\endcsname
  \expandafter\let\csname oldend#1\expandafter\endcsname\csname end#1\endcsname
  \renewenvironment{#1}
     {\linenomath\csname old#1\endcsname}
     {\csname oldend#1\endcsname\endlinenomath}}
\newcommand*\patchBothAmsMathEnvironmentsForLineno[1]{
  \patchAmsMathEnvironmentForLineno{#1}
  \patchAmsMathEnvironmentForLineno{#1*}}
\AtBeginDocument{
\patchBothAmsMathEnvironmentsForLineno{equation}
\patchBothAmsMathEnvironmentsForLineno{align}
\patchBothAmsMathEnvironmentsForLineno{flalign}
\patchBothAmsMathEnvironmentsForLineno{alignat}
\patchBothAmsMathEnvironmentsForLineno{gather}
\patchBothAmsMathEnvironmentsForLineno{multline}
}

\title[Ion Weibel Saturation]{Saturation Level of Ion Weibel Instability and Isotropization Length Scale in Electron-Ion Weibel-Mediated Shocks}

\author[T. Jikei and T. Amano]{
Taiki Jikei,$^{1}$\thanks{E-mail: jikei@eps.s.u-tokyo.ac.jp}
Takanobu Amano,$^{1}$
\\
$^{1}$Department of Earth and Planetary Science, The University of Tokyo, 7-3-1 Hongo, Bunkyo-ku, Tokyo, 113-0033, Japan
}

\date{Accepted 2024 May 1. Received 2024 April 30; in original form 2024 April 1}
\pubyear{2024}

\begin{document}
\label{firstpage}
\pagerange{\pageref{firstpage}--\pageref{lastpage}}
\maketitle

\begin{abstract}
Ion Weibel instability is considered to be the dominant physics for the dissipation in high-Mach number astrophysical shocks such as supernova remnant shocks and gamma-ray burst shocks.
We study the instability dependence on various parameters using theory and particle-in-cell simulations.
We demonstrate that electron physics determines the saturation level of the Weibel-generated magnetic field, even though the instability is driven by the ions.
We discuss the application to astrophysical and laboratory laser experiment environments to clarify the roles of the ion Weibel instability.
We develop a model for the isotropization length scale in Weibel-mediated shocks and compare its value to other characteristic length scales of each system.
We find that electron heating to near equipartition is crucial for the formation of ultra-relativistic Weibel-mediated shocks.
On the other hand, our results imply that non-relativistic shocks in typical interstellar medium are not purely mediated by the Weibel instability. 
\end{abstract}

\begin{keywords}
plasmas -- shock waves -- ISM: supernova remnants -- gamma-ray bursts
\end{keywords}

\section{Introduction} \label{sec:intro}
Astrophysical shocks are considered to be the primary accelerator of cosmic rays because of the well-known efficient shock acceleration mechanisms \citep{Drury1983,Blandford1987}.
In collisional systems, a supersonic flow forms a shock mediated by particle collisions.
However, many astrophysical environments consist of collisionless plasmas. 
Thus, shocks must be mediated by plasma microinstabilities.
Of those collisionless shocks, low \Alfven Mach number $(M_{\mathrm{A}}\lesssim10)$ shocks are relatively well understood \citep{Balogh2013}.
It is known that a fraction of reflected ions from the shock are isotropized via pitch-angle scattering due to ion-cyclotron resonant instabilities \citep{Winske1984,Winske1988}.

It is considered that non-resonant instabilities are dominant in higher Mach numbers.
\citet{Medvedev1999} pointed out that Weibel instability could play an essential role in ultra-relativistic shocks in gamma-ray-bursts (GRBs).
Weibel instability can generate a strong magnetic field, even without a background magnetic field \citep{Weibel1959,Fried1959}.
Therefore, it is the leading candidate for shock mediation in high Mach number shocks, in which the kinetic energy of the upstream plasma is orders of magnitudes larger than the upstream magnetic field energy.
The physics of Weibel-mediated shocks have been studied by theory and particle-in-cell (PIC) simulations both in relativistic \citep{Silva2003,Hededal2004,Frederiksen2004,Medvedev2005,Kato2005,Kato2007,Spitkovsky2008a,Spitkovsky2008b,Bret2013,Groselj2024}, and non-relativistic regimes, which is relevant to supernova remnant (SNR) shocks \citep{Kato2008}.
It has also been studied for laboratory laser plasmas \citep{Ross2012,Fox2013,Huntington2015,Fox2018,Fiuza2020}.

While some studies demonstrate the feasibility of relatively thin Weibel-mediated shocks \citep{Kato2007,Kato2008,Sironi2013,Fiuza2020}, others imply that unmagnetized Weibel shocks should be much thicker.
\citet{Lyubarsky2006} pointed out that the upstream ion gyro radius could be smaller than the Weibel isotropization length scale, even for ultra-relativistic shock parameters, and the GRB shocks may not be mediated by the Weibel instability.
Although some studies ignore the effect of background magnetic field, recent relatively large electron-to-ion mass ratio simulations of non-relativistic shocks also reported the dependence on the background magnetic field strength, even for high Mach numbers, implying these shocks may not be purely mediated by the Weibel instability \citep{Matsumoto2015,Bohdan2021,Jikei2024}.

Although the Weibel instability seems to be a very simple beam instability, it has complicated parameter dependence: beam species (ion or electron), beam velocity, temperatures (both for the beam and the background), the strength of the background magnetic field, etc.
Moreover, typical electron-ion shock simulations use reduced mass ratios to save computational resources.
Because of these complications, a unified understanding of ion Weibel instability for the entire parameter space has not been accomplished despite extensive study for over half a century.
It is not a very good practice to study nonlinear physics and its application without understanding the basic properties of the dominant instability.
This is especially problematic when simulations or experiments cannot use the precise parameters of the target systems, which is almost always the case for astrophysical shocks.
A typical PIC simulation employs a reduced dimension and/or a mass ratio.
Meanwhile, upstream conditions (collisionality, temperature, and magnetization) in a typical laboratory laser experiment differ from those of an astrophysical environment.
The effect of these approximations needs to be understood.

In this paper, we clarify the physics of the ion Weibel instability in an unmagnetized shock.
We start with a theoretical modeling of the growth and saturation level of the Weibel magnetic field.
It will be confirmed by the subsequent 1D and 2D PIC simulations with a wide range of shock velocity, electron temperature, and mass ratios.
We show that electron parameters, rather than the ions, determine the saturation level of the ion Weibel instability in idealized situations where electron screening operates efficiently.
We estimate the isotropization length scale using the idealized simulation results and compare them with astrophysical and laboratory laser experiment parameters.
In many cases, the thickness of a high-Mach number shock is thinner than the Weibel isotropization length scale estimated by the ideal setup.
We show the possible scenarios for shock formation in different parameters, such as relativistic and non-relativistic astrophysical shocks and shocks produced by laser-ablated plasma flows in laboratory experiments.
Results imply that electron parameters, such as temperature and magnetization, are also important in these more complicated systems.

The paper is organized as follows.
The theory of the ion Weibel instability is described in Sec. \ref{sec:Weibel}.
Results of idealized 1D and 2D PIC simulations are shown in Sec. \ref{sec:simulation}.
The isotropization length of ions is estimated, and comparisons to realistic astrophysical shocks and laboratory laser experiments are discussed in Sec. \ref{sec:thickness}.
Finally, a summary and conclusions are given in Sec. \ref{sec:conclusion}.

\section{Theory} \label{sec:Weibel}
\subsection{Model}
The primary motivation of this study is to understand the shock formation in a collisionless unmagnetized electron-ion plasma. To simplify the problem, we consider symmetric counterstreaming cold ion beams in a background of warm electrons described by the Maxwell-J\"uttner (MJ) distribution \citep{Synge1957}:
\begin{equation}
f_{\mathrm{e0}}(\gamma) = n_0
\frac{\gamma \sqrt{\gamma^2-1}}{\theta_{\mathrm{e}} K_2(1/\theta_{\mathrm{e}})}
\exp \left[ - \frac{\gamma}{\theta_{\mathrm{e}}} \right],
\end{equation}
where $\gamma$ is the Lorentz factor of individual particles, $n_0$ is the density, and $\theta_{\mathrm{e}}=k_{\mathrm{B}}T_{\mathrm{e}}/m_{\mathrm{e}}c^2$ is the normalized temperature. 
$K_n$ denotes the $n^{\mathrm{th}}$-order modified Bessel function of the second kind. 
In the following, we work in the electron rest frame, which coincides with the (non-propagating) Weibel magnetic field frame \citep{Pelletier2019,Lemoine2019} in this symmetric setup. 
In other words, the half ($n_0/2$) of the ions are drifting in the positive $x$ direction with the shock (three) velocity $V_{\mathrm{sh}}$ and corresponding Lorentz factor $\gamma_{\mathrm{sh}}=[1-(V_{\mathrm{sh}}/c)^2]^{-1/2}$, and the other half ($n_0/2$) are drifting with the same velocity in the negative $x$ direction.

Notice that while we focus on the shock formation, we consider the three-component plasma with stationary background electrons.
We think this choice is indeed the key to single out the role of electrons in the ion Weibel instability. 
In the typical setup with counterstreaming charge-neutral plasmas, the two electron streams may also give rise to the electron-scale instabilities, which are either electrostatic or electromagnetic in nature, depending on the shock velocity and the electron temperature \citep{Bret2004,Bret2009}.
Particularly, when the system is initially unstable against the electron Weibel instability, the magnetic fluctuations at saturation will be strong enough to partially magnetize the electrons. 
The electron magnetization may affect the response of the electrons to the ion current \citep{Moiseev1963,Lyubarsky2006}, which we will demonstrate plays a significant role in the ion Weibel instability occurring much later in time.
See App.~\ref{A2} for details about the electron-scale instabilities.
In addition, the electron may also experience additional heating through the energy transfer from the ions, which will further affect the electron magnetization, particularly at relativistic temperatures, because of the increase in effective inertia.

To avoid these complications as much as possible, we assume that the electron heating has been completed at an early time to give the prescribed temperature $\theta_{\mathrm{e}}$ before the onset of the ion Weibel instability. 
By doing so, we try to extract the essential physics of the ion Weibel instability under the given electron background. 
We define a parameter $\tau$ for the background electrons by the following relation
\begin{equation}
\theta_{\mathrm{e}}\left[3-\frac{1}{\theta_{\mathrm{e}}}+\frac{K_1(1/\theta_{\mathrm{e}})}{\theta_{\mathrm{e}}K_2(1/\theta_{\mathrm{e}})}\right]=\tau(\gamma_{\mathrm{sh}}-1),
\end{equation}
which characterizes the efficiency of electron heating.

The most natural scenario would be $\tau=1$, which corresponds to the thermalization of the electron flow kinetic energy with a Lorentz factor of $\gamma_{\mathrm{sh}}$ without any energy transfer from the ions. 
This is the standard assumption of electron heating in collisionless shock transition region associated with longitudinal instabilities (e.g., electron two-stream, Buneman, etc.), which results in the elecron thermal velocity of $v_{\mathrm{th,e}}=\sqrt{k_{\mathrm{B}}T_{\mathrm{e}}/m_{\mathrm{e}}}=V_{\mathrm{sh}}/\sqrt{3}$ in the non-relativistic limit. 
We shall call this the fiducial model.

On the other hand, it has been known that a significant fraction of the ion kinetic energy can be transferred to the electrons at relativistic shocks, although with artificially reduced mass ratios \citep{Sironi2013}. 
In our model, the efficiency of energy transfer is parameterized by $\tau > 1$. In the extreme case of $\tau=m_{\mathrm{i}}/m_{\mathrm{e}} \equiv M$, the electron thermal energy is the same as the ion-beam energy. 
We shall call this the equipartition model.
Note that this scenario indicates that the substantial energy transfer from ions to electrons occurs {\it before} the ion Weibel instability sets in, whereas the electron heating in relativistic Weibel-mediated shocks appears to proceed gradually within the Weibel turbulence region \citep{Vanthieghem2022}.
Nonetheless, the equipartition model is relevant to astrophysical scenarios as will be discussed in detail in  Sec. \ref{sec:thickness}.

Using $\tau$, we also define the effective mass ratio $M_{\mathrm{eff}}$.
\begin{equation}
M_{\mathrm{eff}}=\frac{m_{\mathrm{i}}\langle\gamma\rangle_{\mathrm{i}}}{m_{\mathrm{e}}\langle\gamma\rangle_{\mathrm{e}}}=M\frac{\gamma_{\mathrm{sh}}}{\tau(\gamma_{\mathrm{sh}}-1)+1},
\end{equation}
where $\langle\gamma\rangle_s$ is the average Lorentz factor of species $s$.
In the non-relativistic limit $(\gamma_{\mathrm{sh}}-1\ll1)$, $M_{\mathrm{eff}}$ is identical to $M$ regardless of $\tau$.
In the ultra-relativistic limit $(\gamma_{\mathrm{sh}}-1\gg1)$, $M_{\mathrm{eff}}\sim M/\tau$.
Therefore, $M_{\mathrm{eff}} \sim 1$ in the equipartition model $\tau=M$.

Another simplification we will make is to focus on the dynamics of the system transverse to the beam direction. 
This is natural in that the maximum growth of the Weibel instability occurs at the wavenumber perpendicular to the beam. 
Therefore, we naively expect that the saturation can be understood solely by considering the transverse dynamics. 
We shall see that the saturation level in this simplified setup is determined by the single parameter $M_{\mathrm{eff}}$, which is explained by a simple physical argument. 
At first glance, however, it might appear rather inconsistent with the previous studies of the ion Weibel instability, including the longitudinal dynamics along the beam direction \citep{Kato2008,Ross2012}. 
We will discuss the relation between our results and the existing literature, which clarifies the role of the electron heating associated with the longitudinal dynamics (see Sec.~\ref{sec:thickness}).

\subsection{Linear Growth}
Let us now consider the linear phase of the ion Weibel instability.
\citet{Lyubarsky2006,Achterberg2007} derived the growth rate and the saturation level of the relativistic beam, although some approximations concerning the electron distribution have been made.
To the author's knowledge, however, no previous studies explicitly present a corresponding discussion for the non-relativistic case with a finite electron temperature.
This is probably because the importance of electron physics for the non-relativistic ion Weibel instability has been overlooked.
Thus, in the following, we focus on the non-relativistic beam velocity to supplement the theory in the ultra-relativistic limit.

The plasma frequency of species $s$ is defined as $\omega_{\mathrm{p}s}=(4\pi n_0 e^2/m_s)^{1/2}$, where $e$ is the elemental charge and $m_s$ is the mass of each particle species $s$.
Note that this is the {\it non-relativistic} definition for plasma quantities.
Let us also define the {\it relativistic} plasma frequency $\omega_{\mathrm{p}s,{\mathrm{rel}}}=\gamma_{\mathrm{sh}}^{-1/2}\omega_{\mathrm{p}s}$, which will be useful later.
The conductivity tensor $\bm{\sigma}$ defined by $\tilde{\bm{j}}=\bm{\sigma}\cdot\tilde{\bm{E}}$, where $\bm{j}$ and $\bm{E}$ are current density and electric field, respectively.
A Fourier transformed quantity $A$ is denoted by $\tilde{A}$ hereafter.

Taking the wavenumber in $y$ direction $\bm{k} = (0, k, 0)^{\mathsf{T}}$, we obtain $\bm{\sigma}$ as the sum of the following tensors \citep{Stix1992}.
\begin{align}
&\frac{\bm{\sigma}_{\mathrm{i}}}{i\omega}=\frac{\omega^2_{\mathrm{pi}}}{\omega^2}
\mathrm{diag}\left[1+\left(\frac{V_{\mathrm{sh}}k}{\omega}\right)^2,1,1\right], \\
&\frac{\bm{\sigma}_{\mathrm{e}}}{i\omega}=-\frac{\omega^2_{\mathrm{pe}}}{\omega^2}
\mathrm{diag}\left[\zeta_{\mathrm{e}}Z(\zeta_{\mathrm{e}}),1+\zeta_{\mathrm{e}}Z(\zeta_{\mathrm{e}}),\zeta_{\mathrm{e}}Z(\zeta_{\mathrm{e}})\right]. \label{electron}
\end{align}
$Z$ is the plasma dispersion function, whose argument is $\zeta_{\mathrm{e}}=\omega/\sqrt{2}v_{\mathrm{th,e}}k$.
Note that we could use the exact plasma dispersion function for the non-relativistic Maxwell-Boltzman distribution.

The Weibel mode for the cold electron limit $v_{\mathrm{th,e}}=0$ shown below illustrates the role of electrons:
\begin{equation} \label{growthrate}
\frac{\omega}{\omega_{\mathrm{pi}}}=i\frac{V_{\mathrm{sh}}}{c}\left[1+\frac{1}{(\lambda_{\mathrm{se}}k)^2}\right]^{-1/2},
\end{equation}
where $\lambda_{\mathrm{se}}=c/\omega_{\mathrm{pe}}$ is the electron skin depth.
Note that we have made low-frequency $1\ll |ck/\omega|$ and large mass ratio $1\ll M$ approximations.
We can see that the characteristic temporal scale is $\omega_{\mathrm{pi}}$, but the spatial scale is $\lambda_{\mathrm{se}}$.
This indicates that the electron compensates the current by the ions at spatial scales larger than $\sim \lambda_{\mathrm{se}}$.
This is called the electron screening effect \citep{Achterberg2007,Ruyer2015}.

In a realistic situation, the growth time of the ion Weibel instability is slower than the electron response time scale. In this case, a finite electron temperature effect has to be taken into account, as shown by \citet{Lyubarsky2006} for the ultra-relativistic limit. For a hot electron background with $\zeta_{\mathrm{e}} \ll 1$, we can use the small argument expansion of the plasma dispersion function:
\begin{equation}
Z(\zeta_{\mathrm{e}}) \approx -
 i\sqrt{\pi} \left( \frac{\omega_{\mathrm{pe}}}{\omega} \right)^2.
\end{equation}
The growth rate is then given by the solution to the following cubic equation
\begin{equation}
\frac{i\xi}{(\lambda_{\mathrm{se}}k)^3}\left(\frac{\omega}{\omega_{\mathrm{pi}}}\right)^3
-\left(\frac{\omega}{\omega_{\mathrm{pi}}}\right)^2
-\left(\frac{V_{\mathrm{sh}}}{c}\right)^2 = 0,
\end{equation}
where we have defined $\xi=(\pi M/2)^{1/2}(v_{\mathrm{th,e}}/c)^{-1}$. The approximate solution is written as
\begin{equation}
  \frac{\omega}{\omega_{\mathrm{pi}}} \approx
  \begin{cases}
    \displaystyle i
    \xi^{-1/3} \left( \lambda_{\mathrm{se}} k \right)
    \left( \frac{V_{\mathrm{sh}}}{c} \right)^{2/3},
    \quad & k \ll k_{\mathrm{*}},\\
    \displaystyle
    i \frac{V_{\mathrm{sh}}}{c},
    \quad & k \gg k_{\mathrm{*}}. \\
  \end{cases}
\end{equation}
The transition between the long and short wavelength limits occurs at $\lambda_{\mathrm{se}} k_{\mathrm{*}} \approx \xi^{1/3} (V_{\mathrm{sh}}/c)^{1/3} = (\pi M/2)^{1/6}(v_{\mathrm{th,e}}/V_{\mathrm{sh}})^{-1/3}$. 
For the problem of our interest, we have $v_{\mathrm{th,e}}/V_{\mathrm{sh}} \gtrsim 1$, and the transition occurs at a longer wavelength than the cold plasma approximation $\lambda_{\mathrm{se}} k_{\mathrm{*,cold}} \sim1$ (see Eq. \ref{growthrate}). 
In other words, a hot electron background enhances the growth rate at a long wavelength, although the maximum growth rate at a short wavelength remains the same as in the cold plasma case.

The long wavelength shift of the growth rate with a hot electron background is qualitatively understood in terms of the inefficiency of the electron screening effect. 
The condition $\zeta_{\mathrm{e}} \ll 1$ implies that the thermal electrons can easily traverse the instability wavelength over the wave growth time. 
Therefore, the perturbations acting on the majority of electrons cancel, and they are unable to participate in the screening. 
This allows a relatively longer wavelength mode to grow essentially unaffected by the electron screening.

\subsection{Saturation Mechanism}
As we have seen, the electron screening effect is negligible in the short wavelength limit $(\lambda_{\mathrm{se}}k_{\mathrm{*}}\gtrsim1)$. 
This indicates that the instability will be quenched solely by the ion dynamics. \citet{Davidson1972} argues that the Weibel magnetic field saturates when the ion magnetic bounce frequency $\omega_B$ becomes comparable to the growth rate of Weibel instability $\mathrm{Im}(\omega)$.
Let us call this condition the trapping condition.
The bounce frequency is defined as
\begin{equation}
\omega_B=\sqrt{\frac{e}{m_{\mathrm{i}}}k\frac{V_{\mathrm{sh}}}{c}B}.
\end{equation}
Using the asymptotic growth rate $\mathrm{Im} (\omega)/\omega_{\mathrm{pi}} \sim V_{\mathrm{sh}}/c$ in the short wavelength limit $\lambda_{\mathrm{se}} k_{\mathrm{*}} \gtrsim 1$, we can estimate the saturation level as
\begin{equation} \label{trapping}
\varepsilon_{B,\mathrm{trap}} = \frac{B^2}{4\pi m_{\mathrm{i}}n_0V_{\mathrm{sh}}^2} = \frac{1}{4M} (\lambda_{\mathrm{se}} k)^{-2}.
\end{equation}
We have defined beam energy to magnetic field energy conversion rate $\varepsilon_B=B^2/8\pi m_{\mathrm{i}}n_0(\gamma_{\mathrm{sh}}-1)c^2$.
Note that $\gamma_{\mathrm{sh}}-1$ converges to $V^2_{\mathrm{sh}}/2c^2$ in the non-relativistic limit. 
It is easy to understand that the maximum saturation will be achieved at $\lambda_{\mathrm{se}} k_{\mathrm{*}} \sim 1$, where the growth rate deviates from the asymptotic value. 
The exact estimate based on the full numerical solution with a finite electron temperature will be discussed later.
Note that the ions' finite (beam-perpendicular) thermal spread, which we ignore here, reduces the growth rate.
This may also contribute to a lower saturation level if the ion distribution changes in the nonlinear evolution \citep{Lyubarsky2006}.

In contrast, it is natural to anticipate that the electron dynamics also plays an important role in the saturation at long wavelength $(\lambda_{\mathrm{se}}k_{\mathrm{*}}\lesssim1)$ where the electron screening is efficient. 
\citet{Lyubarsky2006} argued that electron magnetization leads to saturation, again in the ultra-relativistic limit. 
We suggest that the same idea holds in the non-relativistic case, indicating that the instability may saturate when the electron gyro-radius becomes comparable to the current filament thickness: $R_{\mathrm{g,e}} = m_{\mathrm{e}}c V_{\mathrm{e}}/eB \sim1/k$ where $V_{\mathrm{e}}$ is the characteristic electron bulk flow velocity.

The above condition may be understood as follows. 
The electrons that participate in the screening are initially accelerated to cancel the ion current to attain a finite drift velocity $V_{\mathrm{e}}$. 
The Lorentz force associated with the perturbation magnetic field on the accelerated electrons tends to evacuate the electrons from the current filament. 
When the Lorentz force becomes significant, the electrons are no longer able to follow the ion motion. 
The evacuation of electrons out of the filament without affecting the ion dynamics will produce a net charge separation across the filament. 
The resulting electrostatic field will then prevent the accumulation of the ion current, thereby quenching the instability. 
By using the initial beam velocity as a characteristic bulk velocity at saturation $V_{\mathrm{e}} \sim V_{\mathrm{sh}}$, we may estimate the saturation level determined by the electron magnetization as follows
\begin{equation}
\varepsilon_{B,\mathrm{mag}}=\frac{1}{M}(\lambda_{\mathrm{se}}k)^2.
\end{equation}
Note that the electron magnetization condition formally resembles the \Alfven current limit \citep{Alfven1939}. 
However, this considers the magnetization of the stable background component rather than the instability-driving population. 
Therefore, the predicted saturation level is smaller than the \Alfven current limit by a factor of $M$.

\begin{figure}
  \centering
  \includegraphics[width=\linewidth]{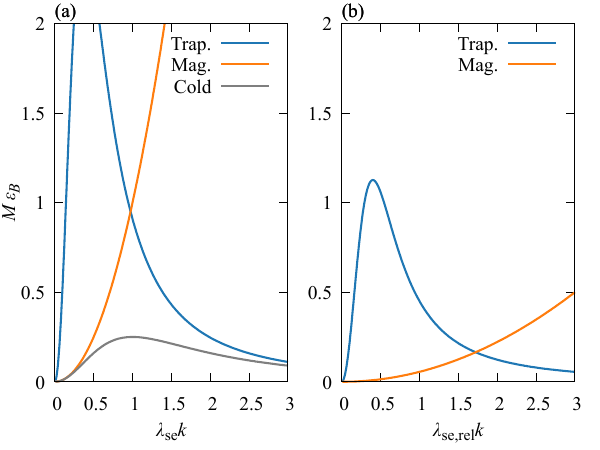}
  \caption{Theoretical estimate of $\varepsilon_B$. Panel (a): the non-relativistic, (b) the relativistic case \citep{Lyubarsky2006}. The blue lines show the trapping, and the orange lines correspond to electron magnetization. The grey line in panel (a) indicates the trapping estimate in the cold electron limit.} 
  \label{Figure1}
\end{figure}

Both estimates based on the trapping and electron magnetization conditions are shown in Figure \ref{Figure1}. Panel (a) is for the non-relativistic limit discussed here with $\tau = 1$ and $(v_{\mathrm{th,e}}=V_{\mathrm{sh}}/\sqrt{3})$, whereas Panel (b) shows the relativistic limit presented by \citet{Lyubarsky2006} with $\tau = 1$. We see that the minimum of the two conditions occurs at the effective electron skin depth scale $1/k \sim \lambda_{\mathrm{se,rel}} = c/\omega_{\mathrm{pe,rel}}$ in both cases. 
It should be noted that the often quoted trapping condition based on the cold plasma approximation (shown in grey) predicts the same order of magnitude estimate both in terms of the saturation level and its spatial scale. 
However, with a finite electron temperature effect (blue), it predicts a much larger saturation level at a longer wavelength, which is not consistent with what we observe in simulations. 
This indicates that electron magnetization (orange) plays a crucial role in the saturation at longer wavelengths.

In summary, the saturation level of the ion Weibel instability is determined by the minimum of the trapping and electron magnetization conditions in both non-relativistic and relativistic regimes. 
The crossover between the two conditions will occur at the effective electron skin depth. 
This gives the simple scaling law of the saturation level:
\begin{equation}
\varepsilon_B \sim \frac{1}{M_{\mathrm{eff}}},
\label{eq:saturation_scaling}
\end{equation}
which we will prove in the following section using PIC simulations. 
Note that this scaling is essentially the same as the one obtained by \citet{Lyubarsky2006} in the ultra-relativistic limit but holds in more general situations.

It is clear from the discussion above and the scaling law that electron physics plays a role in determining the saturation, which may be counter-intuitive in that the instability itself is driven by the ion beams. 
We think that this complication has been the source of various confusion in the literature despite the long history of the Weibel instability since the first discovery \citep{Weibel1959}. 
Of particular note is the fact that the saturation level is dependent on the effective electron temperature when it is highly relativistic because $M_{\mathrm{eff}} \sim M / \tau$. 
This implies that strong electron heating via an efficient ion-to-electron energy transfer will affect the final instability saturation level in the relativistic regime. 
In Sec. \ref{sec:simulation}, we use fully relativistic 1D and 2D particle-in-cell (PIC) simulations to demonstrate that the saturation level does indeed follow the scaling law Eq.~(\ref{eq:saturation_scaling}) under the ideal condition where the electron heating is minimized.

\section{PIC Simulation} \label{sec:simulation}
\subsection{Simulation Setup} \label{subsec:setup}
We perform 1D and 2D PIC simulations with WumingPIC \citep{WumingPIC} to investigate the saturation level of the Weibel-generated magnetic field.
The bulk velocity of the symmetric ion beam is in the $x$ direction.
The simulation is performed in the $y$ direction for 1D and in the $y-z$ plane for 2D.
The background electron is initialized by MJ distribution, which is loaded efficiently by the modified Canfield method \citep{Zenitani2022}.
See Apps. \ref{A1} and \ref{A2} for further discussions regarding the initial condition of the electrons.
We use a grid size of $\Delta x=0.1\lambda_{\mathrm{se}}$ with periodic boundary condition, and a time step of $\Delta t=0.1\omega^{-1}_{\mathrm{pe}}$ for all cases.
We use 5 beam Lorentz factors $\gamma-1=0.01,0.1,1,10,100$ and 4 mass ratios $M=25,100,400,1600$ and let the systems evolve until $\omega_{\mathrm{pi}}t=500$.
Similar setups can be found in previous studies \citep{Ruyer2015,Takamoto2018}.
Other parameters will be described before the result of each model.

\subsection{Fiducial Model}
\subsubsection{1D Simulations} \label{subsec:1D}
First, we present the results for 1D, $\tau=1\,(M_{\mathrm{eff}}=M)$ model.
The simulation grid is in the $y$ direction with a box size of $L_y=112\lambda_{\mathrm{se}}(=1120\Delta x)$.
512 particles per cell (PPC) were used for each species.

\begin{figure}
  \centering
  \includegraphics[width=\linewidth]{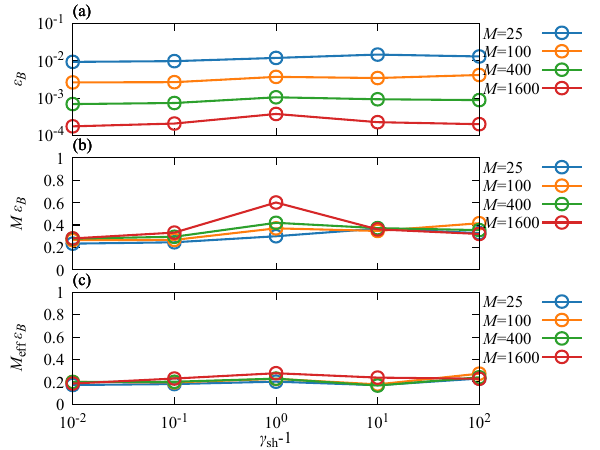}
  \caption{Simulated saturation level $\varepsilon_B$ for 1D $\tau=1$ simulations. Panel (a) shows the $\varepsilon_B$ in logarithmic scale. Panel (b) shows $M\varepsilon_B$ in linear scale. For both panels, blue, orange, green, and red points correspond to $M=25,100,400$ and 1600, respectively. Panel (c) shows the same value with the effective mass ratio $M_{\mathrm{eff}}$.} 
  \label{Figure2}
\end{figure}

Figure \ref{Figure2} shows the spatially averaged beam to magnetic field conversion rate $\varepsilon_B$ using the peak value during the simulation.
Panels (a) and (b) show $\varepsilon_B$ and $M\varepsilon_B$, respectively.
Although we start the simulation from $\tau=1$, i.e., $M_{\mathrm{eff}}|_{t=0}=M$, the electrons gain energy by various processes.
Panel (c) shows $M_{\mathrm{eff}}\varepsilon_B$ using the $\tau$ at the point of time when the magnetic field saturates.
Using this value, we find
\begin{equation} \label{eb}
\varepsilon_B\sim0.2/M_{\mathrm{eff}},
\end{equation}
which is not sensitive to the mass ratio nor the beam Lorentz factor.
This is a result consistent with the theoretical prediction that the saturation level is determined by $M_{\mathrm{eff}}$.
\begin{figure}
  \centering
  \includegraphics[width=\linewidth]{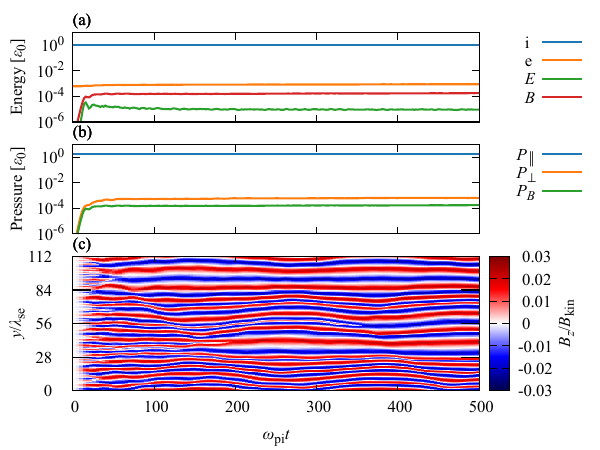}
  \caption{Temporal evolution for $M=1600$, non-relativistic $(\gamma_{\mathrm{sh}}-1=0.1)$ case. Panel (a) shows the average energy density normalized by the initial ion beam energy density $\varepsilon_0$. Blue, orange, green, and red lines correspond to ion, electron, electric, and magnetic fields. Panel (b) shows the average pressure normalized by $\varepsilon_0$. Blue and orange lines correspond to the $x$ and $y$ components of pressure. Green lines show the magnetic pressure. Panel (c) shows the evolution of the Weibel magnetic field $B_z/B_{\mathrm{kin}}$.}  
  \label{Figure3}
\end{figure}

Figure \ref{Figure3} shows the temporal evolution for $M=1600, \gamma_{\mathrm{sh}}-1=0.1$ case.
Panels (a) and (b) show the spatially averaged energy density and pressure evolution, respectively.
These quantities are normalized by kinetic energy density $\varepsilon_0=m_{\mathrm{i}}n_0(\gamma_{\mathrm{sh}}-1)$.
The magnetic field saturates at $\omega_{\mathrm{pi}}t\sim30$.
There is no significant ion and electron energy change throughout the run.
$P_B\lesssim P_{\perp} (=P_y)$ is satisfied in linear and nonlinear evolution.
This indicates that the balance between ion beam-perpendicular pressure and magnetic field determines the spatial profile.
In other words, pressure anisotropy $A=P_{\perp}/P_{\parallel}$ ($=P_y/P_x$ for 1D) is only relaxed to $A\sim\varepsilon_B$ even at quasi-steady state $100<\omega_{\mathrm{pi}}t$.
Panel (c) shows the evolution of the Weibel magnetic field.
Here, the amplitude of the magnetic field is measured in units of $B_{\mathrm{kin}}=[8\pi m_{\mathrm{i}}n_0(\gamma_{\mathrm{sh}}-1)c^2]^{1/2}$.
We see that the initial characteristic spatial scale of the magnetic field is the electron skin depth.
After the saturation of Weibel instability, the spatial scale becomes slightly larger due to filament merging \citep{Vanthieghem2018}.
However, the change of magnetic field energy during this merging phase is negligible compared to the amplification by Weibel instability.

\begin{figure}
  \centering
  \includegraphics[width=\linewidth]{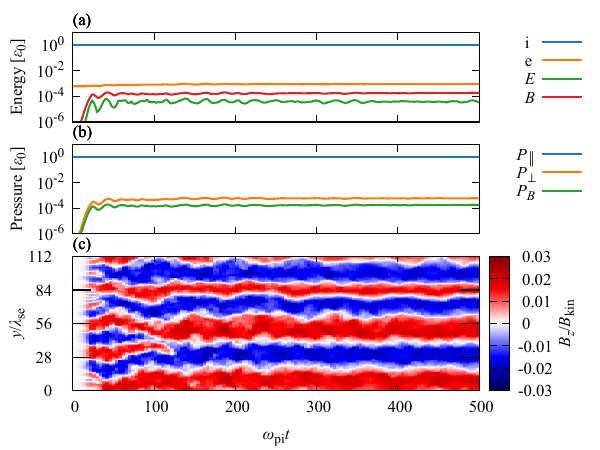}
  \caption{Temporal evolution for 1D, $M=1600$, relativistic $(\gamma_{\mathrm{sh}}-1=10)$ case. The format is the same as in Figure \ref{Figure3}.}  
  \label{Figure4}
\end{figure}

Figure \ref{Figure4} shows the temporal evolution for the $\gamma_{\mathrm{sh}}-1=10$ case.
Energy and pressure evolution (panels (a) and (b)) are mostly similar to the non-relativistic case except for slightly stronger oscillatory behavior in the nonlinear stage.
The magnetic field structure in panel (c) shows a larger spatial scale than the non-relativistic case.
This is because the effective electron skin depth becomes larger $(\sim\lambda_{\mathrm{si,rel}})$ for relativistic cases.
However, the effective mass ratio is still $M$ in the $\tau=1$ model. 
Thus, $\varepsilon_B$ is unaffected by $\gamma_{\mathrm{sh}}$.

\subsubsection{2D Simulations} \label{subsec:2D}
Next, we present the result of 2D out-of-plane beam simulation.
The simulation box is in the $y-z$ plane, where the ion bean is in the $x$-direction.
The box size is $L_y\times L_z=115.2\times115.2 (1152\Delta x\times1152\Delta x)$ with 32 PPC.
\begin{figure}
  \centering
  \includegraphics[width=\linewidth]{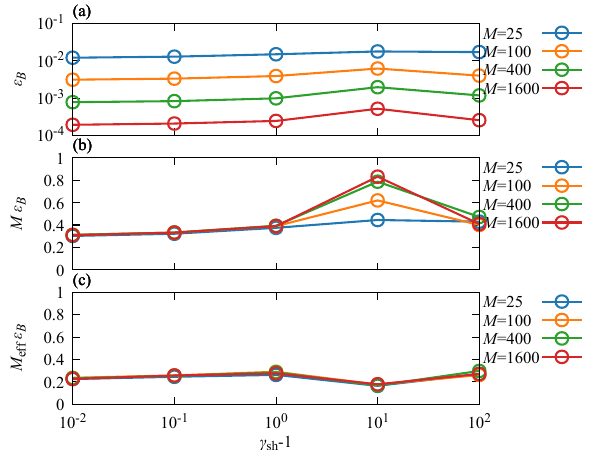}
  \caption{Simulated saturation level for 2D $\tau=1$ simulations. The format is the same as in Figure. \ref{Figure2}.}
  \label{Figure5}
\end{figure}
Figure \ref{Figure5} shows the $\varepsilon_B$ for the 2D case.
The result $\varepsilon_B\sim0.2/M_{\mathrm{eff}}$ is consistent with the 1D case (Figure \ref{Figure2}).

\begin{figure}
  \centering
  \includegraphics[width=\linewidth]{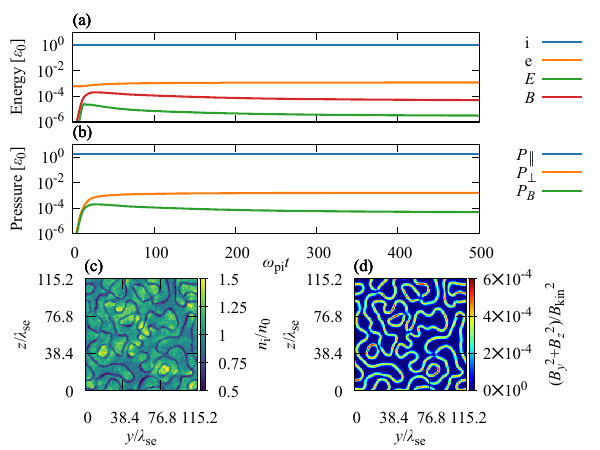}
  \caption{Temporal evolution for 2D, $M=1600$, non-relativistic $(\gamma_{\mathrm{sh}}-1=0.1)$. Panel (a) shows the energy evolution with the same format as in Figure \ref{Figure3} (a). Panel (b) shows the pressure. The format is the same as in Figure \ref{Figure3} (a), but the orange line shows $P_{\perp}(=P_y+P_z)$. Panels (c) and (d) are snapshots of the field taken at $\omega_{\mathrm{pi}}t=100$. (c) shows the ion density normalized by $n_0$. (d) shows the magnetic field energy normalized by $B^2_{\mathrm{kin}}.$}
  \label{Figure6}
\end{figure}
Figure \ref{Figure6} shows the time evolution for the $\gamma_{\mathrm{sh}}-1=0.1$ case.
The evolution of energy and pressure in panels (a) and (b) is consistent with the 1D result (Figure \ref{Figure3} (a) and (b)).
Panels (c) and (d) correspond to the snapshot of ion density and magnetic field energy, illustrating the pressure balance we discussed with the 1D case.

In summary, $\varepsilon_B\sim0.2/M_{\mathrm{eff}}$ regardless of the beam Lorentz factor for the $\tau=1$ model, which is consistent with the theoretical prediction.
Since $A\sim\varepsilon_B$, the Weibel instability alone cannot isotropize the ion beam in the strict sense that the pressure balance is satisfied.
To isotropize the ion and mediate a collisionless shock, the dynamics in the longitudinal direction, which we will discuss in Sec. \ref{sec:thickness}, is crucial.

\subsection{Equipartition Model} \label{subsec:eq}
Finally, we discuss the equipartition $(\tau=M)$ model.
The simulation parameters are $L_y=1152\lambda_{\mathrm{se}}(=11520\Delta x)$ and 512 PPC.
\begin{figure}
  \centering
  \includegraphics[width=\linewidth]{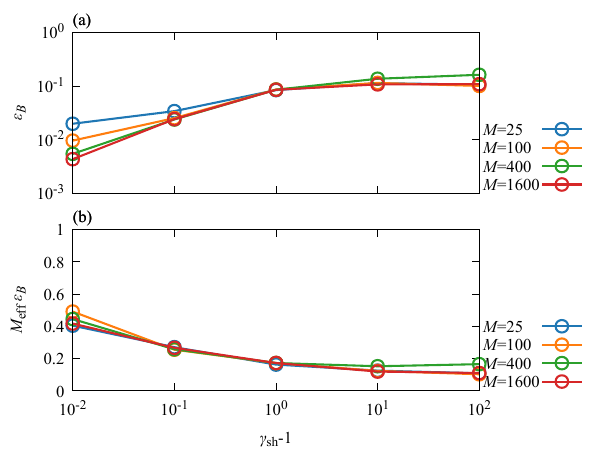}
  \caption{Saturation level for the 1D $\tau=M$ model. The format is the same as in Figure \ref{Figure2}.}
  \label{Figure7}
\end{figure}
Figure \ref{Figure7} (a) shows the $\varepsilon_B$ for the 1D $\tau=M$ model.
In this case, the electrons have the same thermal energy as the protons' kinetic energy.
The mass ratio effects remain in the non-relativistic $(\gamma_{\mathrm{sh}}-1\ll1)$ case.
On the other hand, the effective $M_{\mathrm{eff}}$ asymptotically approach 1 in the ultra-relativistic limit.
The electron screening effect becomes almost negligible in these cases, and energy conversion reaches $(\varepsilon_B\sim0.1-0.2)$.
This value is consistent with the conversion rate of the electron Weibel instability (App. \ref{A1}), in which there is no screening component because the mass of positively charged particles is always comparable to or heavier than the mass of electrons.
Panel (b) shows $M_{\mathrm{eff}}\varepsilon_B$.
The result is consistent with the $\tau=1$ case for relativistic beams $(\gamma_{\mathrm{sh}}-1\gtrsim1)$.
We see up to a factor 2 larger values for non-relativistic cases.
This may be related to the fact that the linear growth rate strongly depends on electron temperature in the non-relativistic regime.
The saturation level by the trapping condition rises because the growth rate increases with higher electron temperature.
Furthermore, the characteristic velocity for the electron magnetization, which we set to $V_{\mathrm{sh}}$ in our theory, may also need some corrections when the electron temperature is extremely large.
On the other hand, the effective mass ratio is the only variable for the relativistic theory.
Also, note that $\gamma_{\mathrm{sh}}-1\sim0.01$ corresponds to $V_{\mathrm{sh}}\sim0.14c$, which is still notably larger than $V_{\mathrm{sh}}\sim0.01c$ expected for SNR shocks.
Since we are mainly interested in shock formation, we do not consider the dependence on other values, such as the beam temperature and number density asymmetry.
However, it should be straightforward to incorporate those effects by changing the growth rate accordingly and using the trapping condition (Eq. (\ref{trapping})).
\begin{figure}
  \centering
  \includegraphics[width=\linewidth]{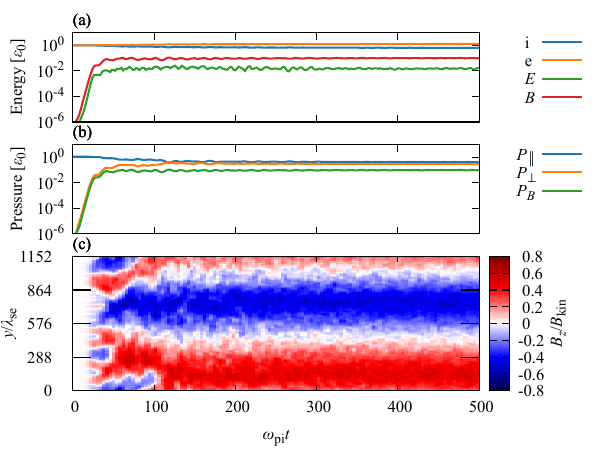}
  \caption{Temporal evolution for $M=1600, \tau=M, \gamma_{\mathrm{sh}}-1=10$. The format is the same as in Figure \ref{Figure3}.}
  \label{Figure8}
\end{figure}
Figure \ref{Figure8} shows the time evolution for the $\gamma_{\mathrm{sh}}-1=10$ case.
Panel (a) shows a larger saturation level due to larger electron energy (i.e., larger effective mass).
The most significant difference from the $\tau=1$ model can be seen in panel (b), in which isotropization $A=1$ is realized at $\omega_{\mathrm{pi}}t\sim150$.
This implies that an ultra-relativistic ion beam can be isotropized, even without the dynamics in the longitudinal direction, within $150\lambda_{\mathrm{si}}$, where $\lambda_{\mathrm{si}}$ is the ion skin depth.

\section{Discussion} \label{sec:thickness}
\subsection{Transverse and Longitudinal Dynamics}
In Secs. \ref{sec:Weibel} and \ref{sec:simulation}, we deliberately limited our discussion to the transverse (beam-perpendicular) dynamics.
This allowed us to isolate the effects of the ion Weibel instability.
As a result, we could obtain a very strong result that the saturation level of the ion Weibel instability is determined only by $M_{\mathrm{eff}}$, which was anticipated by the theory.
Furthermore, $\varepsilon_B\sim0.2/M_{\mathrm{eff}}$ was confirmed for the full range of parameters in using PIC simulations.
In other words, the magnetic field energy can grow up to $\sim0.2$ of the electron energy.
Notably, this value $0.2$ is also consistent with $1/8$ for the cavitation instability driven by a dilute electron-positron beam if the spatial filling factor of cavities is taken into account\citep{Peterson2021, Peterson2022}.
The scenario that the saturation level of magnetic field energy becomes $\sim0.2$ of the electron (or lightest particle species) energy may also apply to any systems in which trapping and magnetization (or \Alfven limit) is the dominant saturation mechanism.

However, the longitudinal (beam-aligned) dynamics may also be crucial for evolution after the saturation of Weibel, including the formation of shocks.
Two major factors that can be dominant for that stage are the kink instability \citep{Daughton1999,Ruyer2018,Takamoto2019} and longitudinal compression.
Kink instability can shorten the longitudinal coherence length of the Weibel filaments.
This introduces a competing process against the stable trapping of the ions, resulting in further isotropization.
It is also natural to assume the compression of the fluid elements (and magnetic field) when advecting from the upstream to downstream can cause extra isotropization.

In principle, these effects can be investigated utilizing 3D PIC simulations.
However, the use of reduced mass ratios is necessary even with modern supercomputers.
Since we found that the effect of the mass ratio is crucial, we would instead choose a theoretical approach of longitudinal modeling based on the transverse results.

\subsection{Isotropization length scale}
Based on the results of the transverse modeling, we make a model for the thickness of Weibel-mediated shocks.
We use a similar approach to previous literature \citep{Lyubarsky2006,Ruyer2016}.
We define the Weibel isotropization length scale $L_{\mathrm{iso}}$ as the thickness needed for an ion beam to be isotropized in (saturated) Weibel turbulence.
Let us assume the following form.
\begin{equation} \label{liso}
\frac{L_{\mathrm{iso}}}{\lambda_{\mathrm{si}}}\sim\max\left[100\left(\frac{M_{\mathrm{eff}}}{20}\right)^{\alpha},50\right].
\end{equation}
$L_{\mathrm{iso}}\sim100\lambda_{\mathrm{si}}$ with $M=20$ was first discovered by \citet{Kato2008}.
The $\varepsilon_B\sim0.01$ in their simulation is also consistent with Eq. (\ref{eb}).
The floor value 50 is inferred from relativistic electron-positron Weibel shock simulations \citep{Kato2007,Sironi2013}.
Here, we introduced a free parameter $\alpha$. 
This parameter describes how the incoming ions interact with the Weibel region.
Longitudinal dynamics, including the kink instability and the shock structure, will determine the value of $\alpha$ \citep{Ruyer2016}.
Note that $\alpha$ could be a universal constant or a function of shock parameters such as $\gamma_{\mathrm{sh}}$.
In the simplest case, in which we do not consider any longitudinal dynamics, $\alpha=1.5$ \citep{Lyubarsky2006}.
On the other hand, \citet{Ruyer2016} proposed $\alpha=0.4$ by a theory using more sophisticated modeling of the shock structure for non-relativistic shocks.
They demonstrate relatively weak mass ratio dependence $(\alpha<1)$ by PIC simulations of up to $M=100$ for full isotropization, and $M=400$ for partial isotropization.
However, $\alpha$ obtained from simulations is slightly larger than $0.4$.
For our purposes, we keep $\alpha$ as a free parameter because we do not know the precise value (or formula) that covers the whole parameter space.

\begin{figure}
  \centering
  \includegraphics[width=\linewidth]{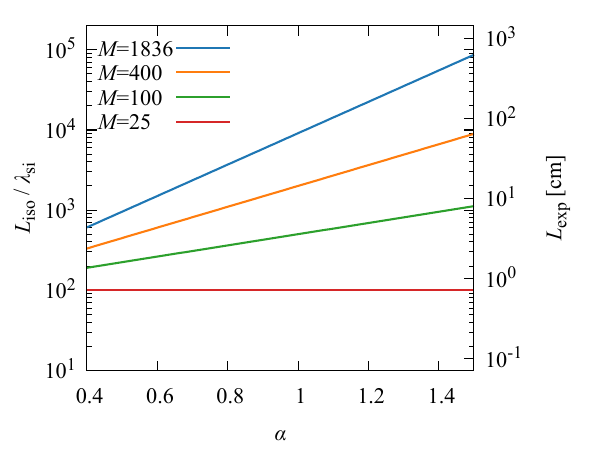}
  \caption{Estimated Weibel isotropization length scale as a function of $\alpha$. Blue, orange, green, and red lines correspond to $M=1836,400,100$, and $25$, respectively.}
  \label{Figure9}
\end{figure}

Figure \ref{Figure9} shows the estimated $L_{\mathrm{iso}}$ as a function of $\alpha$.
$L_{\mathrm{exp}}=L_{\mathrm{iso}}$ calculated with $n_0=10^{19}\,\mathrm{cm}^{-3}$ which is a typical parameter for laboratory laser experiments.
$M=1836, \alpha=1$ predicts $L_{\mathrm{iso}}/\lambda_{\mathrm{si}}\sim10^4$ and $L_{\mathrm{exp}}\sim 100\,\mathrm{cm}$.
These values seem much larger than the values obtained from simulations and experiments.
The same can be said even for $\alpha=0.5$.
We shall discuss below the physics that may be causing this discrepancy.

\subsection{Application to Laboratory Laser Experiments}
We are in the era in which we can study Weibel-mediated plasma beam collision by laser experiments \citep{Fiuza2020}.
A shock was formed within the size of the experiment $\sim3\,\mathrm{cm}$.
This is smaller than the estimated $L_{\mathrm{exp}}$ even for $\alpha=0.5$.
Furthermore, carbons, not protons, are often used in these laser experiments, which makes the spatial scale of the interaction region even larger.
\citet{Grassi2021} argue that one reason for this thin shock in the experiment is the different flow structure.
PIC simulations of astrophysics shocks assume uniform upstreams.
Laser-ablated plasma flow, however, is non-uniform.
They found that this non-uniform flow can form a shock much faster because the fast plasma emitted from the target first generates a strong magnetic field, reflecting the slower ions that enter the interaction region in the late phase.
Other conditions, such as finite collisionality and the existence of a Bierman magnetic field, may also need to be considered in laser experiments \citep{Ross2012,Huntington2015}.

\subsection{Application to Non-Relativistic Astrophysical Shocks}
\begin{figure}
  \centering
  \includegraphics[width=\linewidth]{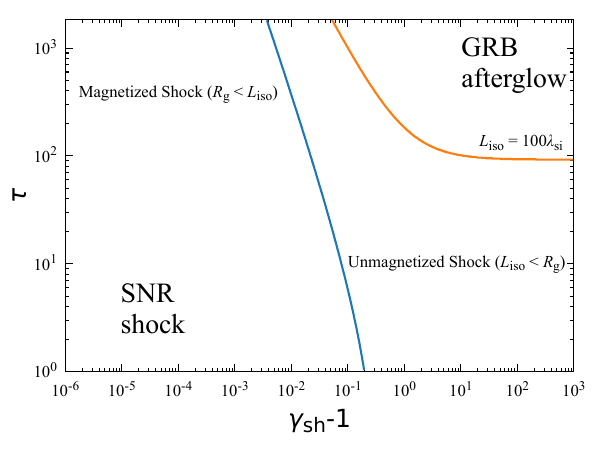}
  \caption{Comparison between ion gyroradius and Weibel isotropization length using typical ISM parameters. The blue line indicates the border between magnetized and unmagnetized shock. The $L_{\mathrm{iso}}=100$ is shown as the orange line.}
  \label{Figure10}
\end{figure}

Now, let us consider astrophysical shocks.
If we naively interpret the result of Fig. \ref{Figure9}, the thickness of high-Mach number shocks are $\sim10^3\lambda_{\mathrm{si}}$ if $\alpha=0.5$ and $\sim10^5\lambda_{\mathrm{si}}$ if $\alpha=1.5$.
On the other hand, the gyroradius of the upstream ions for quasi-perpendicular shocks can be written as
\begin{equation} \label{rg}
\frac{R_{\mathrm{g,i}}}{\lambda_{\mathrm{si}}}=\frac{\sqrt{4\pi mn_0c^2}}{B_0}\sqrt{\gamma_{\mathrm{sh}}^2-1}\sim M_{\mathrm{A}}, 
\end{equation}
Where $M_{\mathrm{A}}$ is the \Alfven Mach number.
In Fig. \ref{Figure10}, we compare $L_{\mathrm{iso}}$ (Eq. \ref{liso}) and $R_{\mathrm{g,i}}$ (Eq. \ref{rg}) using typical interstellar medium (ISM) parameters: $n_0=1\,\mathrm{cm}^{-3}, B_0=10\,\si{\micro G},\alpha=1$.
The blue line shows the border between magnetized $(R_{\mathrm{g,i}}<L_{\mathrm{iso}})$ and unmagnetized $(L_{\mathrm{iso}}<R_{\mathrm{g,i}})$.
Note that this classification relies solely on our shock thickness model.
To confirm this, we may need controlled shock simulations, for instance, that compare quantities like shock formation time, with and without background magnetic field for each parameter $\gamma_{\mathrm{sh}}, B_0$, etc.

The typical velocity of young SNR shocks is $V_{\mathrm{sh}}\sim0.01c\,(\gamma_{\mathrm{sh}}-1\sim10^{-4})$.
This means $R_{\mathrm{g,i}}$ is smaller than $L_{\mathrm{iso}}$.
This holds even for $\alpha=0.5$.
This implies that non-relativistic shocks, such as SNRs (bottom left), are magnetized in the sense that the shock-normal spatial scale is determined by the ion gyration \citep[e.g.,][]{Leroy1982}. 

Furthermore, we recently found another process leading to a larger Weibel magnetic field than the idealized setup.
A finite background beam-perpendicular magnetic field, which magnetizes the electrons, plays a crucial role in non-relativistic $M_{\mathrm{A}}\sim100$ shocks \citep{Jikei2024}.
The magnetic field amplification is enhanced in the presence of a beam-perpendicular magnetic field, which suppresses the electron screening effect \citep{Achterberg2007}.
Furthermore, the electron magnetohydrodynamic (EMHD) dynamo-like process amplifies the beam-aligned magnetic field, which does not exist in unmagnetized systems.
It is considered that the nonlinear dynamics of this beam-aligned anti-parallel field, such as magnetic reconnection, result in efficient electron acceleration \citep{Matsumoto2015,Bohdan2020,Jikei2024}.

\subsection{Application to Relativistic Astrophysical Shocks}
The role of Weibel instability in ultra-relativistic shocks, such as GRB afterglows, is discussed in detail by \citet{Lyubarsky2006}.
Their conclusion is that relativistic shocks in ISM cannot be mediated by the Weibel instability unless the electrons are heated to near equipartition.
They argue that the ion gyration by background magnetic field is not negligible even for ultra-relativistic shocks.
We have confirmed their statements by PIC simulations and longitudinal modeling.
We conclude that the Weibel isotropization length scale $L_{\mathrm{iso}}$ becomes large when the electrons are cold $(\tau\ll M)$, whereas $L_{\mathrm{iso}}\lesssim100\lambda_{\mathrm{si,rel}}$ with near-equipartition $(\tau\sim M)$ electrons.
Note that Figure \ref{Figure10} indicates that ultra-relativistic shocks are unmagnetized shocks even with $\tau\sim1$. 
However, this does not contradict the arguments in \citet{Lyubarsky2006} because the parameters and definition of magnetized/unmagnetized shocks used for Figure \ref{Figure10} are slightly different from theirs.

The physics of relativistic electron-ion shocks has also been studied by open-boundary PIC simulations where the formation of Weibel-mediated shocks has been demonstrated \citep{Spitkovsky2008a,Sironi2013}.
Those results imply that the electrons are heated continuously in the shock transition region.
In other words, electrons are cold $(\tau\sim1)$ near the upstream edge, but they are heated eventually to near equipartition deep inside the shock transition region.
Therefore, the peak saturation level can be as high as $\varepsilon_B\sim0.1$.
The shock thickness in shock simulations is usually defined by the longitudinal length scale in which the large $\varepsilon_B$ is sustained.
The typical value $\sim100\lambda_{\mathrm{si,rel}}$ is fully consistent with our $L_{\mathrm{iso}}|_{\tau\sim M}\lesssim100$.
Theoretical modeling of electron scattering in the Weibel frame also supports this heating scenario \citep{Vanthieghem2022}. 

Finally, an initially ultrarelativistic (unmagnetized) shock could eventually turn into a magnetized shock as it slows down to a non-relativistic speed.
In Figure \ref{Figure10}, the switching point corresponds to $\gamma_{\mathrm{sh}}-1\sim10^{-2}\,(V_{\mathrm{sh}}\sim0.1c)$.
This may be confirmed by investigating the transition from the relativistic phase to the non-relativistic phase of low-luminosity GRB afterglows (typically 10-100 days after the explosion).
\citet{BarniolDuran2015} predicts the light curve with a fixed electron temperature and magnetic field.
However, our results imply that these parameters could change in the long term as the shock decelerates.
The relativistic phase is dominated by unmagnetized Weibel instability in which the electron temperature is near equipartition to the ions.
On the other hand, a magnetized shock is expected in the non-relativistic phase.
This qualitatively different shock structure may result in different characteristics for synchrotron emission.
Incorporating this self-consistent change of afterglow parameters may be essential for a more precise emission model.
Other examples of astrophysical phenomena with $V_{\mathrm{sh}}\sim0.1c$ are ultrafast outflows (UFOs) \citep{King2015} and 
kilonovae \citep{Shibata2019}.
Plasma dynamics and the effect of the magnetic field in these intermediate parameters need to be investigated by future PIC simulations.

\section{Summary and Conclusions} \label{sec:conclusion}
In this paper, we investigated the ion Weibel instability and its applications to laboratory and astrophysical shocks.
First, we discussed the details of the transverse dynamics by theory and PIC simulations.
We have shown that the saturation level of ion Weibel instability can be written as $\varepsilon_B\sim0.2/M_{\mathrm{eff}}$ for the entire range of parameters we have tested.
This is consistent with the theory that the trapping and electron magnetization determine the saturation level.

Then, we used the result of the transverse results to make a model for the formation of Weibel-mediated shocks.
We introduced a parameter $\alpha$ to connect the Weibel field and shock formation and calculated the estimated shock thickness for various systems.

For non-relativistic shocks in laboratory and astrophysical plasmas, the shocks may not be purely mediated by the ion Weibel instability.
Non-uniform upstream and non-negligible collisions may play important roles in laboratory experiments.
Meanwhile, the upstream magnetic field is crucial in weakly magnetized astrophysical shocks such as SNR shocks. 
The gyro-radius of upstream ions is estimated to be smaller than the isotropization length scale by Weibel turbulence.
The physics of ion Weibel instability in these systems and the shock formation need to be further investigated.
Performing a laboratory laser experiment with an external magnetic field may be a good way to obtain better knowledge of SNR shocks.

On the other hand, our results are consistent with previous simulations that imply the ion Weibel instability can generate a very strong magnetic field in ultra-relativistic shocks with hot $(\tau\sim M)$ electrons \citep{Kato2007,Sironi2013}.
Ultra-relativistic shocks such as GRB afterglows can be mediated by unmagnetized Weibel with hot electrons.
However, the mechanism of the intense electron heating needs to be clarified in the future.

Although we mainly discussed ultra-relativistic $(V_{\mathrm{sh}}\sim c)$ and non-relativistic $(V_{\mathrm{sh}}\sim 0.01c)$ shocks, there are many astrophysical phenomena with velocity in between these $(V_{\mathrm{sh}}\sim 0.1c)$.
Our results imply that $V_{\mathrm{sh}}\sim 0.1c$ is where the dynamics of the shock change from unmagnetized to magnetized.
Plasma microphysics in these parameters may also be an important topic for future work.

\section*{Acknowledgments}
This work was supported by JSPS KAKENHI Grants No. 22K03697 and 22J21443. 
T.~J. is grateful to The International Graduate Program for Excellence in Earth-Space Science (IGPEES), University of Tokyo. 
T.~A. thanks for the support from the International Space Science Institute (ISSI) in Bern through ISSI International Team Project \#520 ({\it Energy Partition across Collisionless Shocks}). 
This work used computational resources of the supercomputer Fugaku provided by the RIKEN Center for Computational Science through the HPCI System Research Project (Project ID: hp230073), and the A-KDK computer system at Research Institute for Sustainable Humanosphere, Kyoto University.

\section*{Data Availability}
The data underlying this article will be shared on reasonable request to the corresponding author.

\bibliographystyle{mnras}
\bibliography{reference}

\appendix

\section{Electron Weibel Instability} \label{A1}
To investigate the Weibel instability without screening by oppositely charged components, we discuss the electron Weibel instability in 1D geometry.
We use similar setups as the 1D simulation in Sec. \ref{sec:simulation}.
Electrons are an initially cold counterstreaming component.
The ions are an initially cold background component, and the mass ratio is fixed at $M=1600$.
The simulation length is $500\omega_{\mathrm{pe}}^{-1}$.

\begin{figure}
  \centering
  \includegraphics[width=\linewidth]{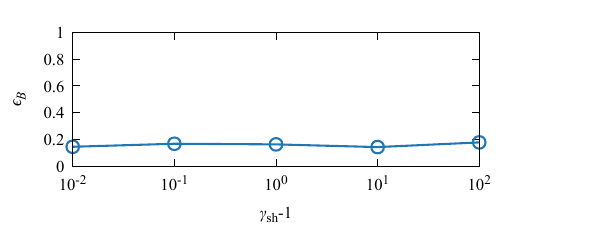}
  \caption{Energy conversion ratio of the electron Weibel instability.}
  \label{FigureA1}
\end{figure}

Figure \ref{FigureA1} shows the electron-beam energy to magnetic field conversion rate.
Let us define $\epsilon_B=B^2/8\pi m_{\mathrm{e}}n_0(\gamma_{\mathrm{sh}}-1)c^2$ and $\epsilon_0=m_{\mathrm{e}}n_0(\gamma_{\mathrm{sh}}-1),$ using lunate epsilon.
$\epsilon_B\sim0.15$ independent of $\gamma_{\mathrm{sh}}$.
This is a similar value to the ion Weibel instability with $M_{\mathrm{eff}}\sim1$.

\begin{figure}
  \centering
  \includegraphics[width=\linewidth]{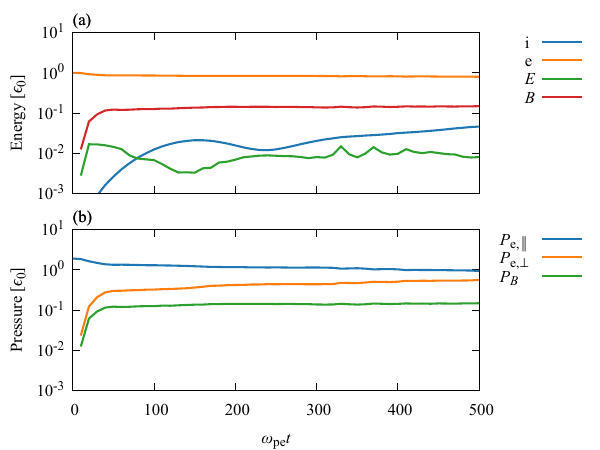}
  \caption{Time evolution for the $\gamma_{\mathrm{sh}}-1=0.1$ case. The format is the same as in Fig. \ref{Figure3} except for normalization, and the plasma pressure components shown in panel (b) are the electron pressure components.}
  \label{FigureA2}
\end{figure}

\begin{figure}
  \centering
  \includegraphics[width=\linewidth]{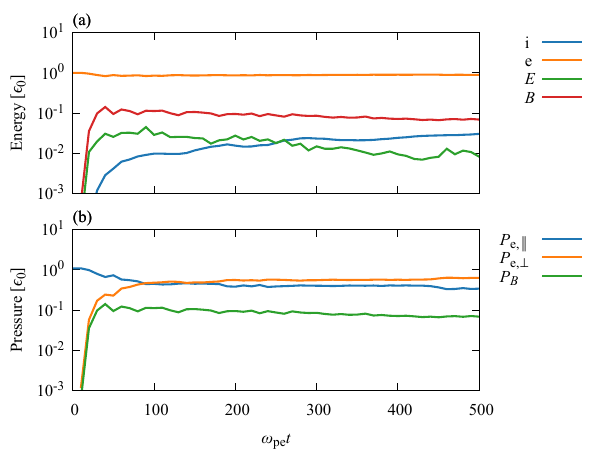}
  \caption{Time evolution for $\gamma_{\mathrm{sh}}-1=10$ case. The format is the same as in Fig. \ref{A2}.}
  \label{FigureA3}
\end{figure}

Figs. \ref{FigureA2} and \ref{FigureA3} show the time evolution for non-relativistic and relativistic cases, respectively.
The time evolution is also similar to the ion Weibel with $M_{\mathrm{eff}}\sim1$.
Electrons are isotropized at $\omega_{\mathrm{pe}}t\sim100\,(\omega_{\mathrm{pi}}t\sim2.5)$, which is shorter than ion Weibel time scale $(\omega_{\mathrm{pi}}T\gtrsim10)$.
However, the setup used in this section is not necessarily realistic because longitudinal, electrostatic modes such as electron two-stream instability are also crucial for an initially cold electron beam distribution.
We discuss the case with both longitudinal and transverse modes in App. \ref{A2}.

\section{Initial Conditions For Electrons} \label{A2}
In this section, we discuss the electron distribution at the early stage, which defines the initial condition for the ion Weibel instability.
We present the 2D in-plane beam simulation.
Note that this is different from the 2D simulation in Subsec. \ref{subsec:2D}, which used out-of-plane beam configuration.
The simulation box is in the $x,y$ plane, where the ion bean is in the $x$-direction.
The box size is $L_x\times L_y=115.2\times115.2(1152\Delta x\times1152\Delta x)$ with 32 PPC, $M=1600$.
Here, both ions and electrons are treated as initially cold beam components.
We investigate the instability within the electron plasma time scale, for instance, electron two-stream instability and electron Weibel instability, by analyzing the electromagnetic field and electron distribution.

For non-relativistic electron beams, we can calculate the growth rate of the electron Weibel instability by the same procedure as the ion Weibel (Sec. \ref{sec:Weibel}).
In the cold plasma limit,
\begin{equation}
\frac{\omega}{\omega_{\mathrm{pe}}}=i\frac{V_{\mathrm{sh}}}{c}\left[1+\frac{1}{(\lambda_{\mathrm{se}}k)^2}\right]^{-1/2}.
\end{equation}
Which has the maximum growth rate of $\sim{\omega_{\mathrm{pe}}V_{\mathrm{sh}}/c}$.
Electron two-stream instability is a longitudinal mode and has the maximum growth rate of $\sim\omega_{\mathrm{pe}}$ \citep{Stix1992}.
For slow beam velocity $(V_{\mathrm{sh}}/c\ll1)$, we expect the two-stream instability to dominate, and for higher velocity, we know that the growth rates of the two modes are comparable.
See \citep{Bret2004,Bret2005} for details including relativistic effects.

\begin{figure}
  \centering
  \includegraphics[width=\linewidth]{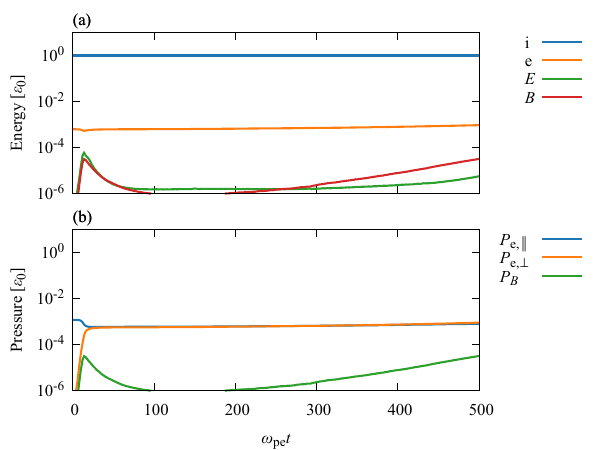}
  \caption{Time evolution of 2D in-plane beam simulation with $\gamma_{\mathrm{sh}}-1=0.1$. The format is the same as in Fig. \ref{Figure3} except the pressure components in panel (b) are the electron pressure.}
  \label{FigureB1}
\end{figure}

Fig. \ref{FigureB1} shows the time evolution for the non-relativistic case.
The electron time scale instabilities isotropize the electrons at $\omega_{\mathrm{pe}}t\sim20\,(\omega_{\mathrm{pi}}t\sim0.5)$.
The electron perpendicular pressure is decoupled from magnetic pressure $(P_{\mathrm{e},\perp}\gg P_B)$ at the onset of ion Weibel.
This implies that the electrostatic modes are more or less responsible for isotropization in non-relativistic parameters.
Note that the velocity used in this simulation is $V_{\mathrm{sh}}/c\sim0.4c$. 
Thus, the growth rate of the electron Weibel instability is relatively high.
The dominance of the electrostatic modes becomes even more significant for slower beam velocity.
When the ion Weibel instability becomes dominant at $\omega_{\mathrm{pe}}t\sim200\,(\omega_{\mathrm{pi}}t\sim5)$. 
The electron energy is almost conserved in this time scale. 
Thus, electrons can be regarded as an isotropic $(\tau\sim1)$ unmagnetized background component for non-relativistic beam cases.

\begin{figure}
  \centering
  \includegraphics[width=\linewidth]{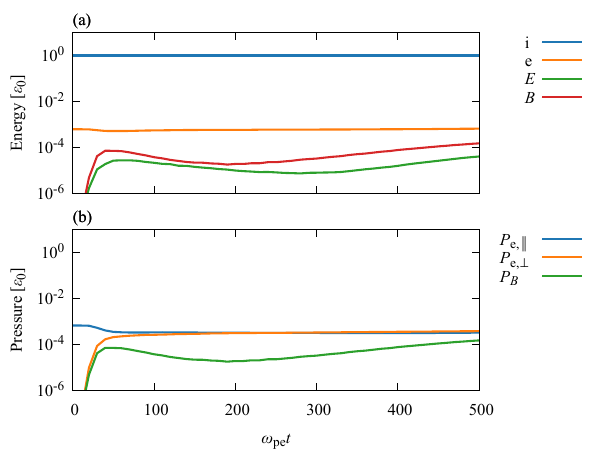}
  \caption{Time evolution for $\gamma_{\mathrm{sh}}-1=10$. The format is the same as in Fig. \ref{FigureB1}.}
  \label{FigureB2}
\end{figure}

Fig. \ref{FigureB2} is the result for the relativistic beam case.
The electrons are isotropized at $\omega_{\mathrm{pe}}t\sim100\,(\omega_{\mathrm{pi}}t\sim2.5)$.
However, contrary to the non-relativistic case, the electron Weibel instability realizes the isotropization.
The electrons are weakly magnetized in the sense that $P_{\mathrm{e},\perp}\sim P_B$ even at the onset of ion Weibel instability.
This magnetization may change the response to the ion Weibel instability.

We have shown that the electrons are isotropized by electron time scale instabilities before the onset of ion Weibel instability for both non-relativistic and relativistic parameters.
For the non-relativistic case, the electrons are not magnetized by electron Weibel because the two-stream instability dominates.
For the ultra-relativistic case, the electrons could be weakly magnetized by electron Weibel instability, which could affect the physics of ion Weibel.
However, previous simulation studies imply that this effect is minor \citep{Ruyer2015,Takamoto2018}.
Furthermore, the electrons in realistic relativistic shocks can be heated to $\tau\gg1$.
In this case, we can also regard the electrons as an unmagnetized background for ion Weibel.

\bsp
\label{lastpage}
\end{document}